\documentclass[aps,prl,twocolumn,groupedaddress,showpacs]{revtex4}

\usepackage{graphicx}
\begin{document}
\title{Nonequilibrium Green's function approach to mesoscopic thermal transport}

\author{Jian-Sheng Wang}
\altaffiliation{Also affiliated with Singapore-MIT Alliance, 
4 Engineering Drive 3, Singapore 117576}
\altaffiliation{and Institute of High Performance Computing, 
1 Science Park Road, Singapore 117528.}
\author{Jian Wang}
\author{Nan Zeng}
\affiliation{Department of Physics, National University of Singapore,
Singapore 117542, Republic of Singapore}

\date{1 May 2006}

\begin{abstract}
We present a formulation of a nonequilibrium Green's function method
for thermal current in nanojunction atomic systems with nonlinear
interactions.  This first-principle approach is applied to the
calculation of the thermal conductance in carbon nanotube junctions.
It is shown that nonlinearity already becomes important at low
temperatures.  Nonlinear interactions greatly suppress phonon
transmission at room temperature.  The peak of thermal conductance is
found to be around 400K, in good agreement with experiments.
High-order phonon scattering processes are important for diffusive
heat transport.
\end{abstract}


\maketitle


Thermal transport in materials has been studied for a long time,
beginning with Joseph Fourier's heat conduction law.  However, a
microscopic theory is possible only after the advent of quantum
mechanics \cite{peierls}.  The earlier treatments are mostly for bulk
systems \cite{rmp-review}.  In recent years, motivated by the
shrinkage of sizes of electronic devices, researchers have paid more
attention to the heat transport phenomena in meso- and nano-scales
\cite{cahill}.  Under such circumstances, some of the concepts have to
be modified.  For example, it has been found that Fourier's law is no
longer valid for many one-dimensional systems \cite{lepri}.  What to
replace it is both interesting theoretically and relevant
experimentally.

A number of approaches have been used to study heat transport, such as
classical molecular dynamics (MD) and the Boltzmann-Peierls equation
\cite{peierls}.  MD can handle nonlinearity, but it is not correct at
low temperatures.  The Boltzmann-Peierls method relies on the concept
of a phonon distribution function in space which is not any more
meaningful in nanojunctions where translational invariance is broken.
The Landauer formula takes care of the low-temperature limit of
ballistic heat transport.  Some attempts have been made to cover both
limits, such as a phenomenological investigation using the concept of
phonon mean free path \cite{wang-jian}.  Recent works in
refs.~\cite{michel,haanggi} are efforts from more fundamental points
of views, starting from quantum principles.  However, these attempts
rely on specific models and approximations.  Clearly, a unified
approach valid for the whole temperature range is still lacking.

In this paper we give a theory for heat transport in nanojunction
using nonequilibrium Green's functions.  Our approach is an exact,
first-principle formulism for general models with nonlinear
interactions, provided that a self-energy can be computed.  This
technique is used extensively in electronic transport
\cite{meir-wingreen}.  Our theory goes beyond linear elastic regime
\cite{ciraci,mingo} by taking nonlinearity perturbatively or through
mean-field approximations.  A comparison of several approximations to
the self-energy on a one-dimensional (1D) chain suggests that
mean-field is reliable up to room temperature.  We then apply the
method to short carbon nanotube junctions and compare with experimental
results.

We consider an insulating solid where only the vibrational degrees of
freedom are important.  The system is composed of a left lead and a
right lead with an arbitrary junction region.  Let the displacement
from some equilibrium position for the $j$-th degree of freedom in the
region $\alpha$ be $u_j^\alpha$, $\alpha = L, C, R$. The quantum
Hamiltonian is given by
\begin{equation}
{\cal H} = \!\!\!\!\!\sum_{\alpha=L,C,R}\!\!\!\!\!H_\alpha  + (u^L)^T V^{LC} u^C + (u^C)^TV^{CR} u^R + H_n,
\end{equation}
where $T$ denotes matrix transpose, $H_{\alpha} = \frac{1}{2}
{(\dot{u}^\alpha)}^T \dot{u}^\alpha + \frac{1}{2} {(u^\alpha)}^T
K^\alpha u^\alpha$, $u^\alpha$ is a column vector consisting of all
the displacement variables in region $\alpha$, and $\dot{u}^\alpha$ is
the corresponding conjugate momentum.  $K^\alpha$ is the spring
constant matrix and $V^{LC}=(V^{CL})^T$ is the coupling matrix of the
left lead to the central region; similarly for $V^{CR}$.  For brevity,
we have set all the atomic masses to 1, but the formulas are equally
applicable to variable masses with a transformation $u_j \to
x_j\sqrt{m_j}$.  Also, we'll set the Planck constant $\hbar$ and
Boltzmann constant $k_B$ to 1.  The nonlinear part of the interaction
is
\begin{equation}
H_n = \frac{1}{3} \sum_{ijk} T_{ijk}\, u_i^C u_j^C u_k^C.
\end{equation}
Quartic interaction can also be handled.

A great simplification is possible due to the linear nature of the
leads $H_L$, $H_R$, and the interaction $V^{C\alpha}$ with the central
region.  Only the central region has nonlinear interactions. The leads
are assumed semi-infinitely long which produce dissipation.  A
traditional approach for many-body problems is to work in second
quantization framework with the phonon creation and annihilation
operators.  Yet, for junction systems without translational
invariance, we find that the notation and the Green's functions will
be simpler if we stay in the first quantization and in the coordinate
representation \cite{doniach}.

The present proposal is parallel to the ideas of the nonequilibrium
Green's function method for electronic transport.  Imagine that at
time $-\infty$ the system is in three separate subsystems in
respective thermal equilibrium at the inverse temperature
$\beta_\alpha$, $\alpha = L, C, R$.  The couplings are switched on
adiabatically, so that at time $t=0$, a steady state is established.
A key starting point is an expression for the heat current.  We begin
with the definition,
\begin{equation}
I_L = - \langle \dot{H}_L(t) \rangle,
\end{equation}
where the decrease in energy of the left lead gives the heat flow to
the central region.  The average is taken with respect to an unknown density
matrix and will be clarified later.  By the Heisenberg equation of
motion, we obtain, at $t=0$, $I_L = \langle (\dot{u}^L)^T V^{LC} u^C
\rangle$.  The expectation value can be expressed in terms of a
Green's function $G^{<}_{CL}(t,t') = - i \langle u^L(t') u^C(t)^T
\rangle^T$. Using the fact that operators $u$ and $\dot{u}$ are
related in Fourier space as $\dot{u}[\omega] = (-i\omega) u[\omega]$,
we get,
\begin{equation}
I_L = - \frac{1}{2\pi} \int_{-\infty}^\infty \!\!\!\!{\rm Tr}\left(
V^{LC} G^{<}_{CL}[\omega]\right) \omega\, d\omega.
\end{equation}
The next step is to eliminate the reference to the lead Green's
functions in favor of the Green's functions of the central region.  We
use the contour ordered Green's function, defined on a Keldysh contour
\cite{keldysh} from $-\infty$ to $+\infty$ and back.  The contour
ordered Green's function can be mapped onto four different normal
Green's functions by $G^{\sigma\sigma'}(t,t') = \lim_{\epsilon \to
0^+} G(t\! +\! i \epsilon \sigma, t'\!+\! i\epsilon \sigma')$, where
$\sigma = \pm (1)$, and $G^{++} = G^{t}$ is the time ordered Green's
function, $G^{--} = G^{\bar{t}}$ is the anti-time ordered Green's
function, $G^{+-} = G^{<}$, and $G^{-+} = G^{>}$.  The retarded
Green's function is given by $G^r = G^t - G^{<}$, and the advanced by
$G^a = G^{<} - G^{\bar{t}}$.  These relations also hold for the self
energy discussed below.  In terms of the contour ordered Green's
function, it can be shown from an equation of motion method or direct
verification by definition, for our model, that $G_{CL}(\tau, \tau') =
\int d\tau'' G_{CC}(\tau, \tau'') V^{CL} g_L(\tau'', \tau')$, where
the integral is along the contour.  The function $g_L$ is the contour
ordered Green's function for the semi-infinite free left lead in
equilibrium at $\beta_L$, e.g., the retarded Green's function in
frequency domain is obtained by the solution of $\bigl[(\omega + i
\eta)^2 - K^L\bigr]g^r_L[\omega] = I$, $\eta \to 0^+$, where $I$ is an
identity matrix.  Using the Langreth theorem and transforming to
Fourier space, the above relation gives us $G^{<}_{CL}[\omega] =
G^r_{CC}[\omega] V^{CL} g^{<}_L[\omega]
+G^{<}_{CC}[\omega]V^{CL}g^{a}_L[\omega]$.  The final expression for
the energy current is
\begin{equation}
I_L = - \frac{1}{2\pi}\!\! \int_{-\infty}^{+\infty}\!\!\!\!\!\!\!\! d\omega\, \omega
\, {\rm Tr}\left( G^r[\omega] \Sigma^{<}_L[\omega] + 
G^{<}[\omega] \Sigma^a_L[\omega] \right),
\label{heat-current}
\end{equation}
where the self-energy due to the interaction with the lead is
$\Sigma_L = V^{CL} g_{L} V^{LC}$.  For notational simplicity, we have
omitted the subscript $C$ on the Green's functions denoting the
central region.

Next, we need a method to compute the full Green's functions.  The
perturbative/diagrammatic expansion is used to derive Dyson equations.
We can treat both the coupling $V^{C\alpha}$ and the nonlinear term
$H_n$ as perturbations, or consider only the nonlinearity as a
perturbation.  The latter is simpler in terms of organization.  The
contour ordered Green's function is expressed in interaction picture:
\begin{eqnarray}
G_{jk}(\tau, \tau') & = & - i \langle T_{\tau} u_j^H(\tau) u_k^H(\tau') \rangle
\nonumber\\
& = & - i \langle T_{\tau} 
u_j^I(\tau) u_k^I(\tau') 
e^{-i \int H_n^I(\tau'')d\tau''}  
\rangle_0,
\end{eqnarray}
where the displacements refer to the central region, the operators in
the top line are in Heisenberg picture; $T_\tau$ is the contour order
operator.  The average $\langle \cdots \rangle_0$ is with respect to
the density matrix of the nonequilibrium steady state when $H_n=0$.
Its explicit form is not known, but the Wick theorem is still valid.
The Green's function $G_0$ of the linear system can be computed from
that of the free subsystems:
\begin{equation}
G_0(\tau,\tau')\! = \! g_C(\tau,\tau')\! +\! 
\int\!\! d\tau_1 d\tau_2 g_C(\tau,\tau_1) \Sigma(\tau_1,\tau_2) G_0(\tau_2,\tau'),
\label{dyson0}
\end{equation}
where $\Sigma = \Sigma_L + \Sigma_R$.  The full nonlinear Green's
function has three types of diagrams in a perturbation expansion.
Diagrams with loops disconnected from the two terminals are zero and
can be dropped.  There is another class of diagrams where the two
terminals are not connected.  Such diagrams are not zero, but are
constants in $\tau$. They represent a thermal expansion effect, and do
not contribute to the heat current in Eq.~(\ref{heat-current}).
Finally, the connected part of the Green's function satisfies a
similar contour ordered Dyson equation relating $G_c$ to $G_0$ using
nonlinear self-energy $\Sigma_n$.  In ordinary Green's functions and
in frequency domain ($\omega$ argument suppressed), the Dyson
equations have solutions \cite{haug}:
\begin{eqnarray}
G_0^r &\!=\!& {G_0^a}^\dagger = \left( (\omega+i\eta)^2 \!-\! K^C 
\!-\! \Sigma^r) \right)^{-1}, \label{G1} \\
G_0^{<} &\!=\!& G_0^r \Sigma^{<} G_0^a, \\
G_c^r &\!=\!& \left( {G_0^r}^{-1} - \Sigma^r_n \right)^{-1}, \\
G_c^{<} &\!=\!& G_c^r \Sigma^{<}_n G_c^a + 
\left( I + G_c^r \Sigma_n^r \right) G_0^{<} \left( I + \Sigma_n^a G_c^a \right).
\label{G4}
\end{eqnarray}
We note that when there is no nonlinearity, $\Sigma_n = 0$, $G =
G_0=G_c$, the heat current formula, Eq.~(\ref{heat-current}), can be
simplified to the Landauer formula,
\begin{equation}
I_L = - I_R = \frac{1}{2\pi} \int_0^\infty \!\!\!d\omega\, \omega\, \tilde{T}[\omega] \left(
f_L - f_R \right),
\end{equation}
with the energy transmission coefficient given by the Caroli formula
\cite{caroli,mingo}, $\tilde T[\omega] = {\rm Tr}\left(G^r \Gamma_L
G^a \Gamma_R \right)$, $\Gamma_\alpha = i(\Sigma_\alpha^r -
\Sigma_\alpha^a)$, where $f_\alpha$ is the Bose-Einstein distribution
function at $\beta_\alpha$.  In order to facilitate comparison with
linear results, we define an effective energy transmission by
$\tilde{T}_{{\rm eff}}[\omega] = \frac{1}{2} {\rm
Tr}\left\{(G^r-G^a)(\Sigma_R^{<}-\Sigma_L^{<}) + i G^{<}
(\Gamma_R-\Gamma_L) \right\}/(f_L -f_R)$.  The function
$\tilde{T}_{{\rm eff}}[\omega]$ is real and even in $\omega$.  Such
effective transmission is temperature-dependent.

\begin{figure}
\includegraphics[width=\columnwidth]{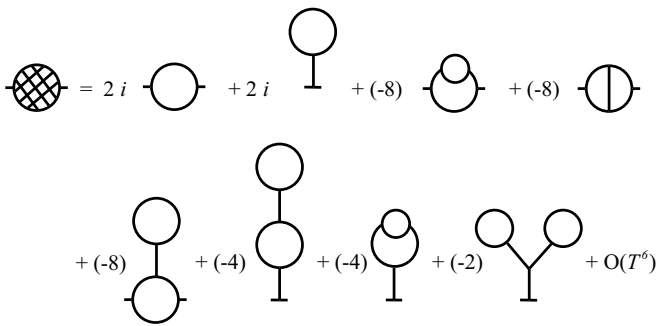}
\caption{\label{fig1}Feynman diagrams for the interaction self-energy
$\Sigma_n$. Each long line corresponds to a propagator
$G_0(\tau,\tau')$; each vertex is associated with the interaction
strength $T_{ijk}$.  All internal site indices are summed and contour
time variables integrated. The number in front of a graph is the
factor multiplying the graph value.}
\end{figure}

So far the equations above are all exact.  Several ways of
approximating the nonlinear contribution to self-energy are possible.
We can simply truncate the diagrams \cite{valle} in the perturbative
expansion for the self-energy.  The diagrams for $\Sigma_n$ to
$O(T_{ijk}^4)$ are shown in Fig.~\ref{fig1}.  These diagrams are still
in the contour variable $\tau$.  For practical calculation, they have
to be changed to real time $t$ in terms of $G^{\sigma \sigma'}$ and
Fourier transformed to the frequency domain.  For example, the leading
order contribution (first two diagrams) to the nonlinear self-energy
is:
\begin{eqnarray}
&&\!\!\!\!\!\!\Sigma^{\sigma\sigma'}_{n,jk}[\omega] \approx 2i 
\!\!\sum_{lmrs}\!\! T_{jlm} T_{rsk} \!\! \int_{-\infty}^{+\infty} 
\!\!\!\!\! G_{0,lr}^{\sigma\sigma'}[\omega']
G_{0,ms}^{\sigma\sigma'}[\omega\! -\! \omega'] 
\frac{d\omega'}{2\pi} + \nonumber\\
&&\;\;\; 2i\sigma \delta_{\sigma,\sigma'} \!\!\!\!\!\!\!\!\!\sum_{lmrs, \sigma''=\pm 1}\!\!\!\!\!\!\!\!\!
\sigma'' T_{jkl} T_{mrs}\!\!
\int_{-\infty}^{+\infty} 
\!\!\!\!\!\!\! G_{0,lm}^{\sigma\sigma''}[0]
G_{0,rs}^{\sigma''\sigma''}[\omega'] \frac{d\omega'}{2\pi}.
\label{Sigma-n}
\end{eqnarray}
Mean-field theory can be obtained by replacing $G_0$ by $G$, and the
equations are solved iteratively.

A general program is implemented based on Eq.~(\ref{heat-current}),
Eq.~(\ref{G1}) to (\ref{G4}), and Eq.~(\ref{Sigma-n}).  The surface
Green's function $g^r_L$ is computed using an efficient recursive
method \cite{surface-green}.  In numerical calculation of the Green's
functions, it is important to keep a small but finite $\eta$, as the
functions are singular in the limit $\eta \to 0^{+}$.  In addition,
on-site potentials are applied to the leads to make the system stable.

\begin{figure}
\includegraphics[width=\columnwidth]{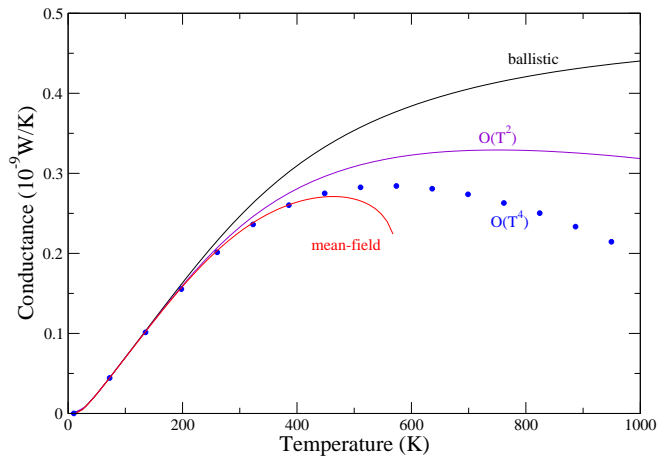}
\caption{\label{fig2}Thermal conductance as a function of temperature
for a 1D junction with three atoms.  The harmonic spring constants are
$k_L = 1.56$, $k_R = 1.44$, $k_C=1.38$ (eV/(\AA$^2$amu)).  The
nonlinear strength is $t=1.8$ eV/(\AA$^3$amu$^{3/2}$).  Small onsite
quadratic potentials are applied to the leads with spring constants
$k_L^{\rm onsite} = 0.01$, and $k_{R}^{\rm onsite} = 0.02$ (eV/(\AA$^2$amu)).}
\end{figure}

We first test various approximations on a 1D junction with parameters
comparable to that of a carbon chain.  The system consists of harmonic
leads and a junction part with harmonic interactions plus cubic
interactions of the form $(1/3) t \sum (u_{j} - u_{j+1})^3$ of
Fermi-Pasta-Ulam type.  Figure~\ref{fig2} presents the results for a
3-atom junction system.  We discuss the effect of nonlinearity to
thermal transport.  As we can see, adding nonlinear contributions
suppresses the thermal conductance at high temperatures.  The
deviation from ballistic transport starts around 200 Kelvin.  As the
temperature rises further, we expect that the higher order graphs
become important.  The high order calculations are rather expensive,
with computational complexity scaled as $O(N^4 M^2)$ where $N$ is
system size and $M$ is sampling points in frequency.  To partially
take into account the higher order contributions but still keep the
computation within reasonable limit for large systems, we find a
mean-field theory is most satisfactory.  In this version of mean-field
treatment, we consider the leading diagrams of $O(T^2)$, and replace
$G_0$ by $G$ only for the first diagram.  The equations are then
solved iteratively.  Good agreement with $O(T^4)$ result is found for
temperatures up to 400 K for the 1D chain.  Thus we expect that the
mean-field theory can give excellent results for moderately high
temperatures.  However, both the perturbative results and mean-field
one break down at sufficiently high temperature, as the cubic
nonlinearity is only metastable.  To predict correctly diffusive
behavior at high temperatures, the quartic interaction and higher
order graphs are essential.

\begin{figure}
\includegraphics[width=\columnwidth]{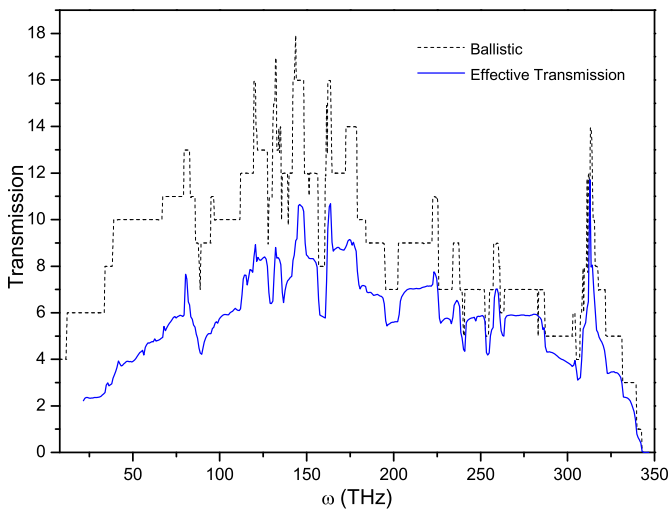}
\caption{\label{fig3} The ballistic transmission and the effective
transmission at 300K for an (8,0) one-unit-cell carbon nanotube
junction.}
\end{figure}

We now turn to the carbon nanotube junctions.  The leads and the
junction are of the same diameter (8,0) nanotube.  The only difference
is that the junction part has cubic nonlinear interactions while the
leads are perfectly harmonic.  The values of $T_{ijk}$ are derived
from the Brenner potential by finite differences.  For computational
efficiency, small values of $T_{ijk}$ are truncated to zero.
Figure~\ref{fig3} shows the ballistic transmission coefficient when
the nonlinearity is set to zero, and is compared with the effective
transmission due to the leading nonlinear effects (the first two
graphs).  It can be seen that the nonlinearity greatly suppresses the
thermal transmission, particularly at low frequencies.  The
discontinuity is smeared out to more smooth curve due to thermal
effect.

\begin{figure}
\includegraphics[width=\columnwidth]{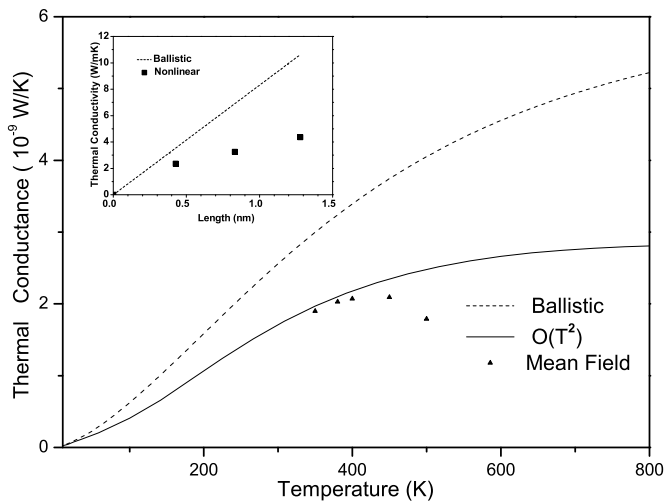}
\caption{\label{fig4} Thermal conductance of (8,0) carbon nanotube
junction with one unit cell (0.426 nm). The inset shows the thermal
conductivity as a function of tube length.  }
\end{figure}

The temperature and length dependence of the nanotube thermal
conductance (conductivity) is shown in Fig.~\ref{fig4}.  The
mean-field theory result gives a peak in the conductance around 400K.
This behavior agrees with experiments \cite{exp-papers} and is in
contrast with MD results which tend to give peaks at much lower
temperatures \cite{md-papers}.  The inset shows the thermal
conductivity calculated from the conductance. The cross-section is
defined as $\pi d^2/4$, where $d$ is the diameter of the tube.  If we
assume that the transport up to 1 $\mu$m is still quasi-ballistic,
then we can estimate that the thermal conductivity at the experimental
accessible length is about 2000 W/(mK).  This is qualitatively in
agreement with experimental values \cite{exp-papers}.

We proposed a fully quantum mechanical approach for computing heat
current of solid junctions with nonlinear interactions.  We have
demonstrated the method with 1D and molecular junction systems.  The
perturbation expansion for self-energy works well up to room
temperatures.  However, it is still a challenge to find efficient and
good approximation for the self-energy at high temperatures.

This work is supported in part by a Faculty Research Grant of the
National University of Singapore.

\bibliography{basename of .bib file}

\end{document}